# Band topology and non-trivial surface States in type-II Dirac semi-metal X(Ni, Pd)Te$_2$

N. K. Karn[1,2*] and V.P.S. Awana[1,2]

[1]Academy of Scientific & Innovative Research (AcSIR), Ghaziabad, 201002, India
[2]CSIR- National Physical Laboratory, New Delhi, 110012, India

**Abstract:**

The semi-metals having electrons near the Fermi level follow the relativistic equation of motion, and show Dirac or Weyl-type behavior. Their orbital resolved electronic bands analysis indicates the non-trivial topological states. Through the first principle methods, the electronic band structure of bulk and thin layers of transition metal dichalcogenides NiTe$_2$ and PdTe$_2$ are calculated and analyzed. Implementation of Hubbard corrections to the band structure indicates zero magnetization, though atoms of Ni are magnetic. Z2 invariant index is calculated for topological classification for both materials and is found to possess strong topology indicating robust topological surface states. Further, Fermi surfaces and topological surface states are computed into k$_z$ plane showing the presence of Dirac-fermions by forming of Dirac cone for PdTe$_2$ but for NiTe$_2$ we do not get clear evidence.

*Corresponding Author
Navneet kumar Karn: E-mail: nkk15ms097@gmail.com
Senior Research Fellow
CSIR-NPL, New Delhi, India, 110060
Ph. +91-11-45609357

**Introduction:**

In the past decade, the quest for quantum materials with novel properties have been on a high and continues till today. Among these, materials with topological properties are mostly studied [1-4]. Topological insulators have a bulk bandgap; however, the gap is nonequivalent to traditional insulator vacuum due to surface state bands [5]. Apart from topological insulators, topological semimetals(TSM) has drawn the attention of condensed matter physicist as it provides an alternate platform to study the relativistic Dirac fermions [1, 3-5]. TSMs possess not only conducting non-trivial topological surface states but their bulk is also conducting, unlike the topological insulators. The semi-metals having electrons near the Fermi level follow a relativistic equation of motion, and show Dirac or Weyl-type behavior depending on the degeneracy of bands [6, 7].

Among various TSMs, the Transition Metal Dichalcogenides(TMD) are of special interest, as it is a 2-D layered material in which atomic layers are separated by a Vander Waals gap. This makes them a good candidate for material intercalation [7, 8]. TMDs like $PtTe_2$ [9], $PtSe_2$ [10], $PdTe_2$ [8,11-14], and $NiTe_2$ [15-19] are considered to be type-II Dirac semimetal. Among these, $PdTe_2$ is found to be superconducting at 1.7K [20], by Au intercalation the $T_c$ can be increased up to 4.7K [8]. Topological property with superconductivity makes $PdTe_2$ a potential material to realize topological superconductivity. The observed MR in these TMDs is not linear and is found to increase in a quadratic manner [14, 15]. The band structure analysis of these TMDs reveal that the Dirac point in these materials lies far below the Fermi level [12, 16, 18]. So, the transport in these materials is dominated by non-relativistic carriers. The Ni and Pd counterpart of these materials, such as $NiTe_2$ and $PdTe_2$ has been extensively studied in the context of topological properties [11-19]. These intrinsic noble characteristics of TSMs, including the topological superconductivity and non-linear MR, increase the eagerness to study the surface states.

Topological materials are characterized based on the symmetry they possess and associated invariants to that symmetry. For systems respecting time-reversal symmetry, the topology is characterized by Z2 invariant index. There have been only a few scattered reports of Z2 of these two TMD [16]. However, here for the first time we provide a topological classification of the two TMDs in terms of Z2 invariant calculated by Wannier Charge Center (WCC) method. We report, both TMDs show strong topological character, with non-trivial surface states. Also, the surface states are calculated from the iterative greens function method from the different planes (110). The structure of these two TMDs is generated in VESTA [21] and we report the simulated XRD peaks of the same. To incorporate the van der Waal interaction between the 2D layers of the two TMDs, DFT-d3 corrections are implemented.

**Computational Methods:**

All computational simulation is performed on open-source software Quantum Espresso [22, 23] which works on the first principle-based DFT. The crystal structure is optimized and self-consistent functional calculations (SCF) have been performed to obtain the ground-state electron density and wavefunction. The electronic exchange and correlation corrections are incorporated by using the Perdew-Burke-Ernzerhof (PBE) type generalized gradient approximation (GGA) functionals [24]. The wave functions were expanded in a plane wave with a charge cut-off of 360 units and wave function cut-off of 45 Ry on a Monkhorst-Pack k-grid of 13×13×8. This computation produces the bulk electronic band structure and projected Density of states (PDOS) of both TMDs. To find any effect of magnetism on bands, DFT+U (Hubbard correction) calculations are implemented. Further, the Bloch wave functions generated by DFT are wannierized using WANNIER90 software [25] which produces maximally localized Wannier functions (MLWFs). The electronic band structure is recalculated using MLWFs and matched with the result obtained by the direct DFT method to check the quality of obtained MLWFs. Moreover, an effective tight-binding model is obtained based on these MLWFs for the material under investigation. For topological properties and calculation of Z2 invariant, the tight-binding model is post-processed in the software WANNIER TOOLS [26]. The Wannier charge centers (WCC) are obtained of whose evolution in Brillouin zone planes indicated the states of Z2-invariant. The surface spectral function was calculated using the iterative surface Green's function method [27, 28] along the (001) plane.

**Results and Discussion:**

Both compounds under investigation $PdTe_2$ and $NiTe_2$ crystallize in trigonal with the P-3m1(164) space group which belongs to centrosymmetric space groups. The structure is a 2D layered structure, which is separated by Van Der Waal's gap. The Waykoff position of the atoms X(Ni, Pd) is 1a (0, 0, 0) and Te atoms at 2d (0.33, 0.66, 0.25). The structure generated using VESTA software is shown in fig. 1(a) and 1(b). Cell parameters of the $NiTe_2$ unit cell are a = b = 3.855 Å & c = 5.277 Å, α = β = 90.0, and γ=120.0; and the same of $PdTe_2$ are a = b = 7.631 Å & c = 9.717 Å, α = β = 90.0 and γ =120. Theoretically, XRD patterns are also calculated using VESTA as shown in figure 1(c) confirming the used phase in the calculation is correct. The XRD peaks also match with the reported XRD peaks of $NiTe_2$ [29] and $PdTe_2$ [14].

*Band Structure:*

The theoretical calculation is performed using the DFT-based software Quantum espresso. The high symmetric path in the first Brillouin zone(BZ) is chosen for band structure calculation as shown in the inset of fig1(c). This is the optimized path suggested by the work of Hinuma et. al. [30]. Since both materials belong to the same crystal class, their first BZ is similar having the same high symmetric points. The suggested path is A-H-L-A-Γ-M-K-Γ. The calculated band structure of $NiTe_2$ is shown in fig. 2(a) where as for $PdTe2$ the same is shown in figure 2(d). It has been found that higher levels of theory, such as DFT-D3 [31], MP2 [32], and B3LYP[33], give more accurate results. Since the crystal under study has van der Waals gaps between its layers, we also included the van der Waals correction, for which DFT-D3 calculation was performed [31]. The dispersion correction energy was evaluated as − 0.83 eV. Both TMDs show very similar band structure features. In both cases, without SOC bands show a double degenerate Dirac-cone in between the path L-A around 0.5eV above the Fermi energy. However, when SOC parameters are included the band degeneracy is lifted. Another triply degenerate point is observable along path A-Γ which decomposes into one four-fold degenerate point and a doubly degenerate band. Fig. 2(b) and (e) show the orbital projected density of states (PDOS) without SOC while figs 2(c) and (f) show the PDOS when SOC is included. From the PDOS, we know the orbital contributions for the bands near the Fermi level. In both cases, the d orbital of transition metal and the p orbital of Te atoms. The total DOS near the Fermi level is finite, which confirms the semi-metallic nature of both TMDs. Also, note that the Dirac cone is classified as type I or II based on its tilt. Here in the band structure, both Dirac cones above discussed are tilted, therefore both TMDs are regarded as type-II Dirac semimetals.

It is well known that Ni has a magnetic moment, so we performed magnetization calculations including the Hubbard corrections. We observe that the net magnetization is zero for both TMDs. Therefore, the implementation of Hubbard corrections to the band structure does not alter the electronic band structure. This implies that Ni atoms in the unit cell have opposite magnetization i.e. atoms are in an antiferromagnetic state. Thus we have a non-magnetic system. The system respects the time-reversal symmetry(TRS) which is confirmed through the SOC band structure calculation. If TRS is not followed, the double degeneracy would have been lifted but here each band is doubly degenerate.

Further, to compute Z2 invariant and surface states, the wavefunctions from first principle calculations are wannierized and the band analysis confirms the orbital contributions for the near Fermi-level bands. From these wannierized we have constructed tight-binding Hamiltonian and implemented it in Wannier Tools. Using the iterative Green's function methods, the surface state spectrum of the two TMDs is calculated. The surface card is set to be in the plane (110), and the calculation was done by taking 81 slices of one reciprocal vector. The path taken for surface state spectrum calculation is K→ Γ →K.

The surface state spectrum of NiTe$_2$ is shown in fig. 3(a) and the same of PdTe$_2$ is shown in fig. 3(b). We observe that the Dirac cone is present at the Γ point around 0.5 eV above the Fermi level in the case of PdTe$_2$. However, the same feature is not sharply present in the case of NiTe$_2$. Thus, the computational simulation indicates the presence of surface states in the 110 planes in PdTe$_2$ but the same is not clearly evidenced in the case of NiTe$_2$. In the next section, the calculation of the Z2 invariant confirms that both are equally likely to possess a non-trivial topological surface state.

*Z2 – invariant:*

Based on the symmetry principle, there are several topological invariants to classify the topology present in the system. For the systems respecting TRS, Z2 invariant [34]; for TRS broken phase chern number; for crystals with reflection symmetry mirror chern number [35]; for topological magnetic phases, Z4 [36] are some of the topological invariants following the symmetry present in the crystal. Here, we find that the TMDs under study are non-magnetic and respect TRS, so the Z2 invariant is to be computed for its topological classification. Different Z2-states are separated by a topological phase transition which involves an adiabatic transformation of the bands leading to the gap closing of the two bands. There are several methods to calculate Z2 invariant - discrete method involves finding Pfaffians/parity over the fixed points of the time-reversal symmetry[34], Fukui- Hatsugai method [37], and by tracking the windings of the Wannier Charge Center (WCC) evolution around the BZ [38, 39]. Here, we use the WCC method to compute the Z2 invariant which is incorporated into WANNIER TOOLS. The WCC evolved by changing the k vector, odd or even number of exchange of these charge centers in different k-planes, defining non-trivial or trivial topology respectively. So from the tight-binding model obtained for Wannier functions, the WCCs are computed numerically and evolved in the 6-planes – $k_x$, $k_y$, $k_z = 0$ and $k_x$, $k_y$, $k_z = 0.5$ in the Brillouin zone. An odd number of the crossing of WCC implies a topologically non-trivial state (Z2=1), whereas an even number of crossings indicate the presence of a topologically trivial state (Z2=0). The computed Z2 invariants for both TMDs are shown in table 1.

For 3D systems, the Z2 topological invariants are represented by four indices ($\upsilon_0$; $\upsilon_1$ $\upsilon_2$ $\upsilon_3$). These indices are calculated by the formula [38, 39]

$$\upsilon_0 = \sum \text{ of } Z2 \in k_i \text{ plane mod } 2$$
$$\upsilon_i = Z2 \text{ value} \in (k_i = 0.5) \text{ plane}$$

where i=x,y,z. The first index designates a strong topology and the rest three indices indicate a weak topology present in the system. Without SOC bands we encounter degeneracies but with SOC bands we do not encounter any band degeneracy. This allows us to compute the Z2 invariant for SOC bands. From Table 1, following the above method, we get the Z2 index for both TMDs as (1; 000) which is exactly the same as the previously well-established topological insulator Bi$_2$Te$_3$ [40] with a strong topological index. The weak topological index shows trivial states at the surfaces $k_i = 0.5$ for both TMDs. However, both TMDs demonstrate strong non-trivial topological phases with $\upsilon_0 = 1$. This gives clear evidence of band inversion when SOC comes into play as degeneracies are lifted. From the surface states calculation compared to the Z2 invariant obtained we observe non-trivial topological surface states are more prominent in PdTe$_2$ compared to NiTe$_2$.

**Conclusion:**

To summarize, in the present work, through the first principle methods, the electronic band structure of bulk and thin layers of transition metal dichalcogenides NiTe2 and PdTe2 are calculated and analyzed. Both belong to the same class of compounds which crystallizes trigonal lattice with P-3m1(164) space group which is centrosymmetric. The projected density of states calculation confirms the semi-metallic nature as it turns out to be finite. This also shows the valance and conduction band near the Fermi level are mostly originating from the d orbital of transition metal and the p orbital of Te atoms. Magnetic calculation shows the net magnetic moment is zero for both systems though one contains highly magnetic Ni atoms. Therefore, the implementation of Hubbard corrections to the band structure does not alter the electronic band structure. The electronic band dispersion in $k_z$ plane shows the formation of the Dirac cone at the center of $k_z$ plane. With no magnetization, band structure analysis tells that the high symmetric point Γ is a time-reversal invariant momentum point. This allows us to calculate Z2 invariant index for both systems for topological classification and is found to possess strong topology indicating robust topological surface states. Further, Fermi surfaces and topological surface states are computed into $k_z$ plane of the reciprocal space showing the presence of Dirac - Fermion by the formation of Dirac cone for PdTe$_2$ but for NiTe$_2$ we do not get clear evidence.

**Acknowledgment:**

The authors would like to thank Director NPL for his keen interest and encouragement. N.K. Karn would like to thank CSIR for the research fellowship. N. K. Karn would also like to thank AcSIR for Ph.D. registration.

**Conflict of Interest Statement:**

The authors have no conflict of interest.

**Figure Captions:**

**Figure 1:** Trigonal unit cell of (a)NiTe$_2$ (b) PdTe$_2$ with space group P-3m1. (c)The simulated XRD peaks of the two TMDs. The inset figure shows the first Brillouin zone, with the Green arrows directing the path chosen for the Band structure calculation.

**Figure 2:** (a) The electronic band structure of NiTe$_2$ with and without SOC parameters including DFT-D3 corrections. (b) and (c) show the projected density of states without and with SOC respectively for NiTe$_2$. (d) Shows the electronic band structure of PdTe$_2$ with and without SOC parameters. (e) and (f) show the projected density of states without and with SOC respectively for PdTe$_2$.

**Figure 3:** This shows the surface state spectrum, (a) for NiTe$_2$ a band degeneracy is there at Γ but the Dirac cone is not apparent. (b) surface state spectrum PdTe$_2$. Dirac cone is observable at Γ point around energy 0.5 eV.

**Figures:**

Figure 1.

(a)  (b)

(c)

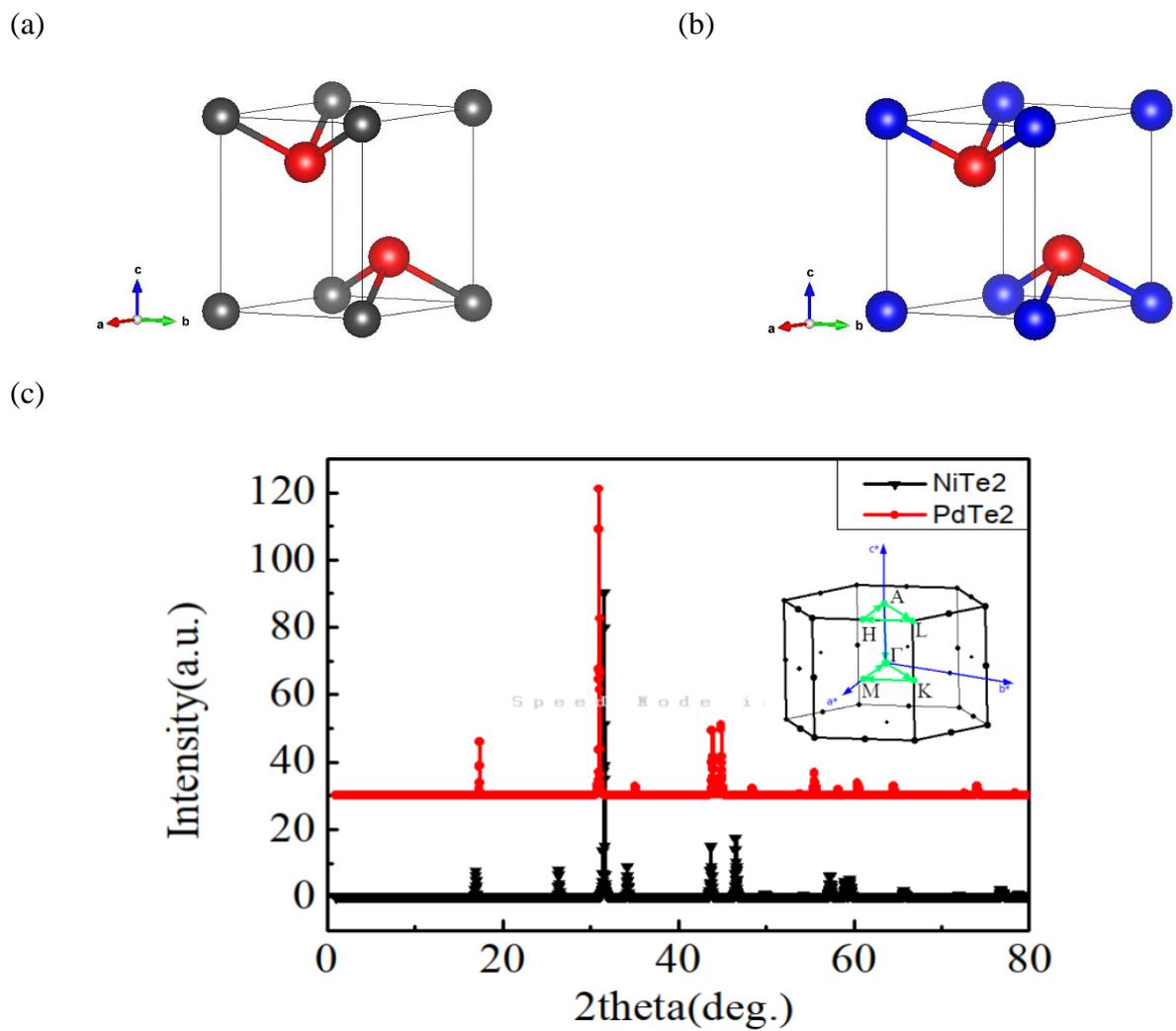

**Figure 2**

(a)  (b)  (c)

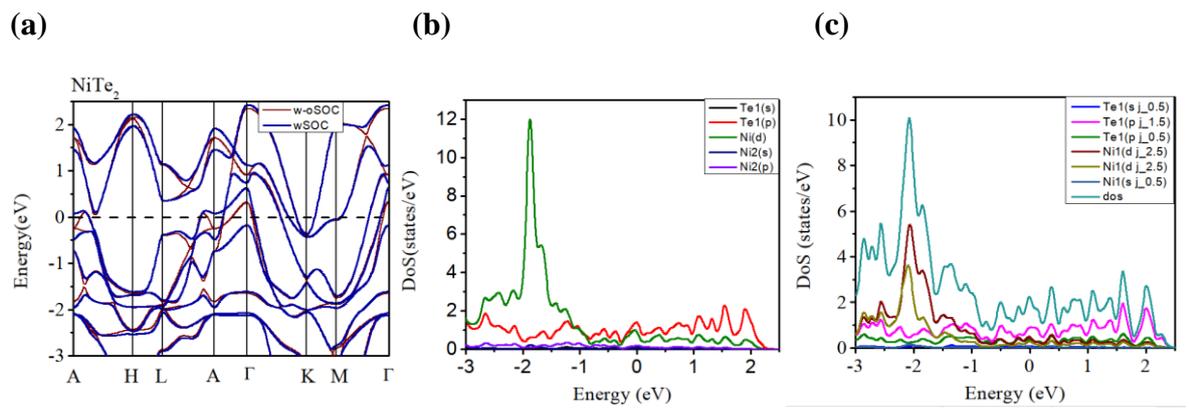

**(d)** **(e)** **(f)**

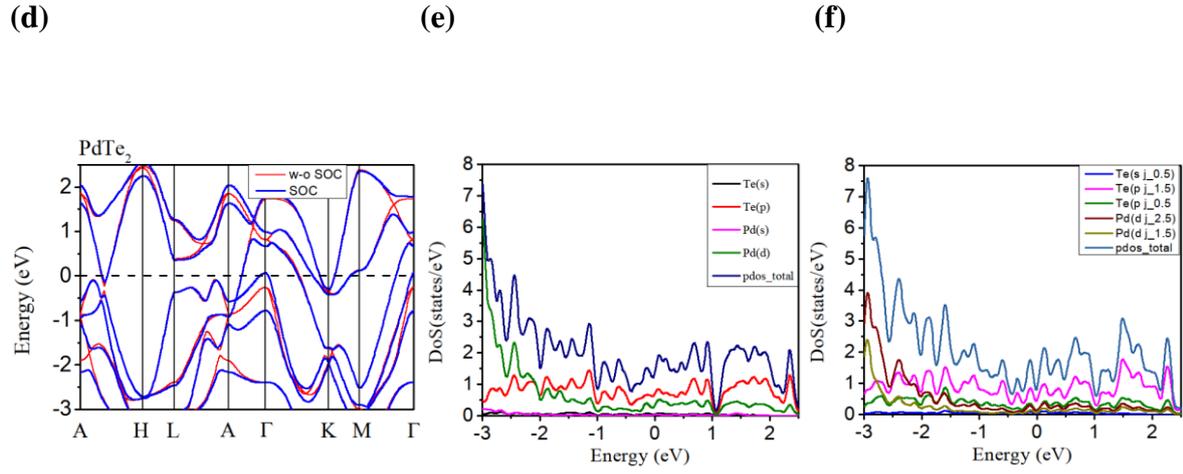

**Figure 3.**

**(a)** **(b)**

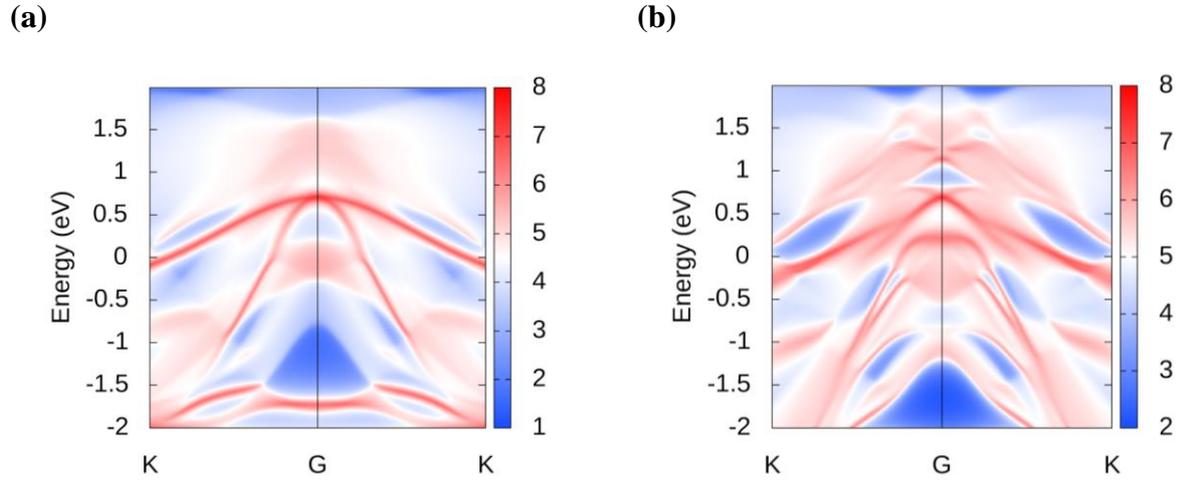

**Table 1**

**Z2 invariant**

| Plane | NiTe$_2$ | PdTe$_2$ |
|---|---|---|
| k$_x$ = 0.0, k$_y$–k$_z$ plane | 1 | 1 |
| k$_x$ = 0.0, k$_y$–k$_z$ plane | 0 | 0 |
| k$_x$ = 0.0, k$_y$–k$_z$ plane | 1 | 1 |
| k$_x$ = 0.0, k$_y$–k$_z$ plane | 0 | 0 |
| k$_x$ = 0.0, k$_y$–k$_z$ plane | 1 | 1 |
| k$_x$ = 0.0, k$_y$–k$_z$ plane | 0 | 0 |